%% file: paper_v6.tex
\documentclass[usenatbib,usegraphicx]{mn2e}
\usepackage{bm}
\usepackage{morefloats}
\usepackage{hyperref}
\usepackage{tabularx}
\usepackage{graphicx}
\usepackage{multirow}
\usepackage{pdflscape}
\usepackage{amsmath}

\title[Mean proper motions and internal velocity dispersions of Galactic globular clusters]
{Mean proper motions, space orbits and velocity dispersion profiles of Galactic globular clusters derived from {\it Gaia} DR2 data}
\author[Baumgardt, Hilker, Sollima \& Bellini]{H. Baumgardt$^{1}$\thanks{E-mail:
h.baumgardt@uq.edu.au}, M. Hilker$^{2}$, A. Sollima$^{3}$, A. Bellini$^{4}$\\
$^{1}$ School of Mathematics and Physics, The University of Queensland, St. Lucia, QLD 4072, Australia \\
$^{2}$ European Southern Observatory, Karl-Schwarzschild-Str. 2, 85748 Garching, Germany\\
$^{3}$ INAF Osservatorio Astronomico di Bologna, via Gobetti 93/3, Bologna, 40129, Italy\\ 
$^{4}$ Space Telescope Science Institute, 3700 San Martin Drive, Baltimore, MD 21218, USA\\
}

\begin{document}

\date{Accepted 2018 xx xx. Received 2018 xx xx; in original form 2018 xx xx}

\pagerange{\pageref{firstpage}--\pageref{lastpage}} \pubyear{201x}

\maketitle

\label{firstpage}

\begin{abstract}
We have derived the mean proper motions and space velocities of 154 Galactic globular clusters and the velocity dispersion profiles of 141 globular clusters based on a
combination of \textit{Gaia} DR2 proper motions with ground-based line-of-sight velocities. Combining the velocity dispersion profiles derived here with new measurements
of the internal mass functions allows us to model the internal kinematics of 144 clusters, more than 90\% of the currently known Galactic globular cluster population.
We also derive the initial cluster masses by calculating the cluster orbits backwards in 
time applying suitable recipes to account for mass loss and dynamical friction. We find a correlation between the stellar mass function of a globular cluster and 
the amount of mass lost from the cluster, pointing to dynamical evolution as one of the mechanisms shaping the mass function of stars in clusters.
The mass functions also show strong evidence that globular clusters started with a bottom-light initial mass function.
Our simulations show that the currently surviving globular cluster population has lost about 80\% of its mass since the time of formation. 
If globular clusters started from a log-normal mass function, we estimate that the Milky Way contained about 500 globular 
clusters initially, with a combined mass of about $2.5 \cdot 10^8$ M$_\odot$. For a power-law initial mass function, the initial mass in globular clusters could have been a factor of three higher.
\end{abstract}


\section{Introduction} \label{sec:intro}

The second data release of the \textit{Gaia} mission \citep{gaiadr2main2018} has provided five astrometric parameters for more than 1.3 billion stars in the Milky Way. 
Compared to its predecessor {\tt HIPPARCOS} \citep{perrymanetal1997}, the median accuracy in the proper motions and parallaxes has increased by a factor
50 while the number of studied stars has increased by more than a factor 10,000. In addition, \textit{Gaia} DR2 contains line-of-sight velocities for more than 7 million stars. 
With this wealth of information of unprecedented accuracy, \textit{Gaia} DR2 has a profound impact in 
many fields of astrophysics, like stellar astrophysics \citep{babusiauxetal2018}, the study of the dynamics of the Milky Way and other nearby galaxies \citep[e.g.][]{fritzetal2018}
and even the study of massive black holes in the distant universe \citep[e.g.][]{wolfetal2018}.

In the present paper we apply the \textit{Gaia} DR2 data to study the motion and internal kinematics of globular clusters in the Milky Way. 
\citet{helmietal2018} have already determined the mean proper motions of about
half of all Milky Way globular clusters based on the \textit{Gaia} DR2 data, while \citet{vasiliev2018} has determined the mean proper motion of 75 of the remaining clusters. In this paper we add several 
additional clusters which so far have no determined
proper motion, resulting in a near complete sample of globular clusters with measured proper motions. In addition, we also improve the accuracy of the mean line-of-sight velocities of globular clusters 
by using the \textit{Gaia} proper motions to remove field stars from the sample of stars with measured line-of-sight velocities.
The main part of the paper is then devoted to study the internal kinematics of the globular clusters, deriving proper motion dispersions from the \textit{Gaia}a DR2 data and combining these with stellar
line-of-sight velocities to determine the internal velocity dispersion profiles. We then use $N$-body models to derive the present-day masses of the clusters.

We thus derive a complete picture of the Galactic globular cluster population in terms of present-day orbital and structural parameters. This information will
be useful to understand the formation history of the Milky Way halo (accreted vs. in-situ) \citep[e.g][]{myeongetal2018}, the total mass of the Milky Way and the shape of its dark matter halo using globular cluster
orbits \citep[e.g][]{postihelmi2018}, the formation of massive star clusters in the early universe, and to understand the evolution and 
mass loss of globular clusters and the contribution of stars lost by them to the bulge and stellar halo of the Milky Way \citep[e.g][]{brandtkocsis2015,fragioneetal2018}, and we will briefly discuss some of these questions
in our paper. 

Our paper is organized as follows:
In section 2 we describe the selection of cluster members in the \textit{Gaia} catalogue, the derivation of the mean proper motions and the proper motion velocity dispersion profiles 
and the $N$-body fits to the observational data. Sec. 3 presents our results and in sec. 4 we draw our conclusions.

\section{Analysis}

\subsection{Cluster sample and selection of target stars}

Our target list of globular clusters was taken from the 2010 edition of \citet{harris1996}, which lists 157 globular clusters. To this list we added the following three clusters
that have been found to be Globular clusters:  Crater \citep{laevensetal2014}, FSR 1716 \citep{minnitietal2017} and Mercer~5 \citep{merceretal2005,longmoreetal2011}.
We omitted the clusters Koposov~1 and Koposov~2 since \citet{paustetal2014} found that they are more likely to be several Gyr old open clusters removed from the Sagittarius dwarf galaxy. 
In addition, they found that both clusters have masses of less than 2,000~M$_\odot$, but steep stellar mass functions, implying that Koposov~1 and Koposov~2 also formed as low-mass objects
and are not stripped down versions of globular clusters.
We also removed GLIMPSE-C01 for which \citet{daviesetal2011} concluded that it is a 400-800 Myr, intermediate-age disc cluster, and BH 176 which was also found
to be an old, metal-rich open cluster that could belong to the galactic thick disc \citep{davoustetal2011, sharinaetal2014}.

We are thus left with a sample of 156 globular clusters. For each of these clusters we selected all {\it Gaia} stars within a circle of 1200 arcsec around
the cluster centers given by either \citet{goldsburyetal2013} or \citet{harris1996}. For IC~1257 and Ter~10 we found that the cluster centers given in \citet{harris1996} were incorrect and determined
new centers from the positions of the member stars found in the \textit{Gaia} DR2 catalogue. For each globular cluster, we then cross-correlated the \textit{Gaia} positions 
against the catalogue of 40,000 globular stars with line-of-sight velocity measurements compiled by \citet{bh18} from published literature and archival ESO and Keck spectra. 
Using a search radius of 1'', we were able to find \textit{Gaia} DR2 counterparts for about 95\% of all stars with measured line-of-sight velocities from \citet{bh18}.
For each cluster star, the individual line-of-sight velocity measurements from the ESO/Keck Science Archives, published literature measurements and \textit{Gaia} DR2  
were averaged and if found too discrepant from each other, the stars were flagged as possible binaries and not used in the later analysis. If necessary, the \textit{Gaia} line-of-sight velocity measurements were  
shifted in each cluster to bring them onto a common mean velocity with the \citet{bh18} velocities. The necessary shifts were usually less than 0.5 km/sec, in agreement
with the systematic offset in the \textit{Gaia} line-of-sight velocities found by \citet{katzetal2018} for faint stars. 


\subsection{Derivation of the mean proper motions}

From the sample of stars found within 1200 arcsec of the center of each cluster, we removed all stars that had 
line-of-sight velocities which deviated by more than $3\sigma$ from the mean cluster velocity. We also removed stars with parallaxes that deviate by more than $3\sigma$ from
the expected parallax calculated using the distance given in \citet{harris1996}. We also removed stars with \textit{Gaia} $G$ magnitudes brighter than 1 magnitude
or $B-R$ colors redder than 0.5 magnitudes than the tip of the RGB to exclude 
obvious non-members from the determination of the mean cluster motion and internal velocity dispersion. To guide us in choosing these limits,
we calculated MESA isochrones \citep{dotter2016,paxtonetal2015} for each individual cluster and varied the assumed extinction until we obtained the
best match of the MESA isochrone against the position of the cluster members
in a $G$ vs. $B-R$ color-magnitude diagram. We also removed all stars which showed significant astrometric excess
noise $\epsilon_i>1$ together with an astrometric excess noise significance $D>2$ as given in fields 33 and 34 of the final \textit{Gaia} catalogue, or for which the astrometric goodness-of-fit parameter given in the \textit{Gaia} DR2 catalogue was larger than 3.5.  
\begin{figure*}
\begin{center}
\includegraphics[width=\textwidth]{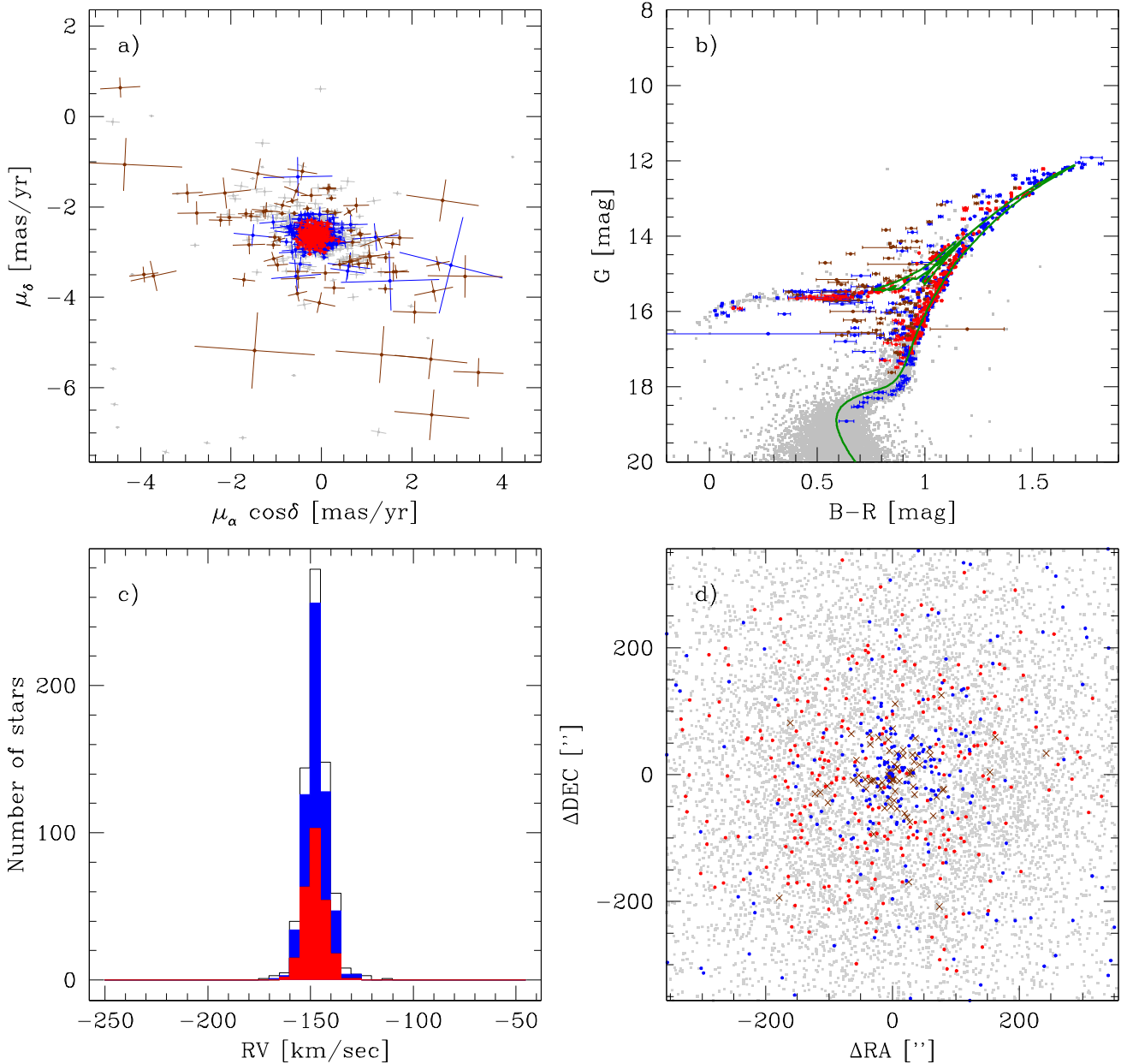}
\end{center}
\caption{Selection of cluster members for the globular cluster NGC~5272 (M3). In all panels stars with measured line-of-sight velocities that were used for the determination of the
mean cluster motion and the internal velocity dispersion profile are shown in red, stars with line-of-sight velocities that were discarded due to e.g. too large proper
motion errors are shown in blue and stars that are members based on their line-of-sight velocities but non-members according to their proper motions are shown in brown. Stars without 
line-of-sight velocities but having a matching proper motion are
shown in grey. Panel a) depicts the proper motion distribution of all stars in the field of NGC 5272, panel b) shows a color-magnitude diagram
of NGC 5272 with the best-fitting MESA isochrone shown by a green line, panel c) shows the line-of-sight velocity distribution of stars and panel d) depicts the spatial distribution
of stars in a 600'' by 600'' field around the cluster centre. The proper motions of the individual stars were rotated into a coordinate system in which both components are statistically independent of each other.}
\label{fig:clex}
\end{figure*}

We furthermore removed stars which had
too large proper motion errors. The maximum velocity error up to which we accepted stars was varied from cluster to cluster, but was generally of the order of about
1.5 times the central velocity dispersion of the giant stars, i.e. about 5 to 15 km/sec. We found that accepting stars with too large proper motion errors or astrometric excess noise 
generally lead to an overestimation of the proper motion dispersion profiles, however for small enough cut-offs the dispersion profiles did not depend on the chosen cut-off.
For a few distant halo clusters or for heavily obscured bulge clusters, which had only few members in the \textit{Gaia} catalogue, we accepted stars with
velocity errors up to 50 km/sec for the calculation of the mean cluster motion.
We also applied cuts in distance from the cluster centre to minimize field star contamination and excluded the centres of a few bright clusters where the stars showed
large deviations from the mean motion of the cluster. The
outer distance limits were varied on a cluster to cluster basis, but were usually of the order of 100 to 700 arcsec, depending on the field star density and how well 
the cluster proper motion separates a cluster from the field stars. We adopted about twice as large distance limits for stars with line-of-sight velocity measurements
than for stars without line-of-sight velocity information, since a matching line-of-sight velocity significantly reduces the probability that a cluster star is a field star.
Fig.~\ref{fig:clex} illustrates our selection process of cluster members for the globular cluster NGC~5272 (M~3).

Our first estimate for the mean proper motion of each cluster was then calculated by averaging the mean proper motion of all stars that are known cluster members
based on their line-of-sight velocities. For clusters located in fields with a very strong background density of stars, we obtained a first estimate by visually inspecting the proper motion distribution 
of all stars within 50 arcsec of the cluster centre. We then determined all stars that have a proper motion within 2$\sigma$ of the mean cluster motion, taking into account
the intrinsic velocity dispersion of the cluster as well as the proper motion error of each star, and calculated a new estimate for the mean proper motion of a cluster based on these possible members. 
This process was repeated until the mean cluster motion and the list of cluster members became stable between successive iterations. 

The mean proper motion, its error and the correlation between the two proper motion components were determined from the member stars as follows: Starting
from eq.~10 of \citet{lindegrenetal2000}, the combined log Likelihood for an ensemble of $i$ stars each with its own proper motion $(\mu_{\alpha* i}, \mu_{\delta i})$ ,
associated errors $\sigma_{\mu_{\alpha* i}}$, $\sigma_{\mu_{\delta i}}$ and covariance $\rho_i$ is given by:
\begin{equation}
\begin{aligned}
\ln L = & -\sum_i \frac{1}{(1-\rho_i^2)} \frac{(\mu_{\alpha* i}-<\!\mu_\alpha*\!>)^2}{\sigma^2_{\mu\alpha* i}+\sigma^2_{Cl}} \\
 &  - \sum_i \frac{1}{(1-\rho_i^2)} \frac{(\mu_{\delta i}-<\!\mu_\delta\!>)^2}{\sigma^2_{\mu\delta i}+\sigma^2_{Cl}}  \\
 &  + \sum_i \frac{2\rho_i}{1-\rho_i^2} \frac{(\mu_{\alpha* i}-<\!\mu_\alpha*\!>)(\mu_{\delta i}-<\!\mu_\delta\!>)}{\sqrt{\sigma^2_{\mu\alpha* i}+\sigma^2_{Cl}}\sqrt{\sigma^2_{\mu\delta i}+\sigma^2_{Cl}}} \;\; .
\end{aligned}
\end{equation}
Here $<\!\mu_\alpha*\!>$ and $<\!\mu_\delta\!>$ are the mean cluster motions in right ascension and declination and $\sigma_{Cl}$ is the internal velocity dispersion of the cluster, which we either
took from the best-fitting $N$-body models of \citet{bh18} or from our new fits that we derived in this paper. Calculating the first derivative of the
above equation with respect to $<\!\mu_\alpha*\!>$ and $<\!\mu_\delta\!>$ and setting it to zero gives the best-fitting values for the mean motion. The associated errors and the correlation between
the proper motion components were then calculated from the second derivatives according to:
\begin{eqnarray}
\nonumber \sigma_{<\mu_\alpha*>} & = & \frac{1}{\sqrt{\frac{\partial^2 ln L}{\partial <\mu_\alpha*>^2}}} \\
\nonumber \sigma_{<\mu_\delta>} & = & \frac{1}{\sqrt{\frac{\partial^2 ln L}{\partial <\mu_\delta>^2}}} \\
 \rho_{<\mu_\alpha* \mu_\delta>} & = & \frac{1}{\sqrt{\frac{\partial^2 ln L}{\partial <\mu_\alpha*> <\mu_\delta>}}} \;\;\; .
\end{eqnarray}

\begin{figure}
\begin{center}
\includegraphics[width=\columnwidth]{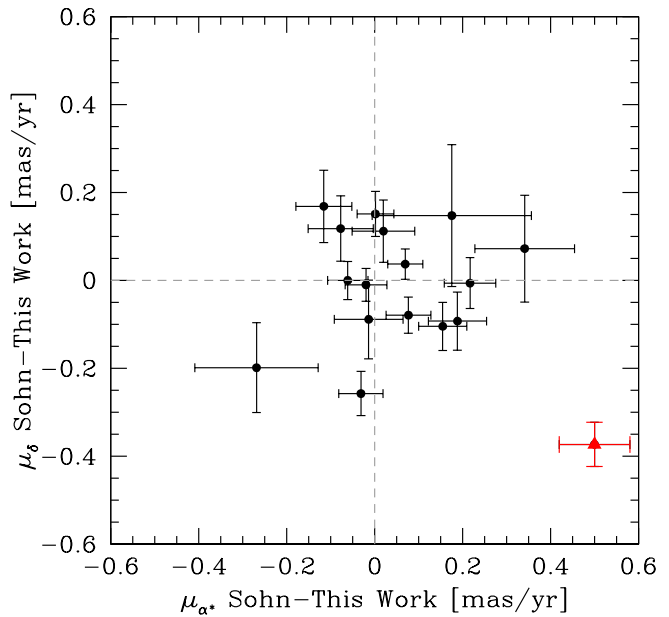}
\end{center}
\caption{Comparison of the mean cluster proper motions derived here with the ones derived by \citet{sohnetal2018} from \textit{HST} data. The differences between the proper motions
are larger than expected based on the formal error bars. Most discrepant is the globular cluster NGC~5466 (shown by a red triangle), for which the proper motion difference is about 
5 times larger in both directions than the combined error estimates.}
\label{compsohn}
\end{figure}

The mean proper motions that we have derived are in good agreement with the ones found by the \citet{helmietal2018} and \citet{vasiliev2018}, however since their data is also based on the
\textit{Gaia} DR2 data, their proper motions do not provide a fully independent check of our results.
Fig.~\ref{compsohn} shows the differences in the mean proper motions derived in this work vs. the mean cluster proper motions that were derived by \citet{sohnetal2018} from multi-epoch \textit{HST} data.
It can be seen that the resulting residuals are larger than expected based on the statistical errors alone since the proper motions of some clusters deviate by more than 3 $\sigma$
from each other. 
Most discrepant is NGC~5466, for which the \textit{Gaia} and \textit{HST} proper motions differ by more than 5$\sigma$ in each direction, pointing either to
strong systematics in the \textit{Gaia} catalogue in the field of NGC~5466 or a problem in the \textit{HST} data. 
Excluding NGC~5466, the residuals in the \textit{HST} vs. the \textit{Gaia} proper motions are in agreement within the uncertainties if we assume a systematic error of 0.10 mas/yr in either the
\textit{Gaia} or \textit{HST} proper motions, in good agreement with the value of 0.07 mas/yr derived by \citet{lindegrenetal2018} for the size of systematics on small angular scales in the \textit{Gaia} DR2 proper motions.

Table~\ref{tab:meanpm} presents the mean cluster proper motions that we derived from the \textit{Gaia} DR2 data. For GLIMPSE~C02 and 2MASS-GC01 we could not reliably identify
any cluster members in \textit{Gaia} DR2 and so these clusters are not listed. For the other clusters we derived mean proper motions based on between 2 and 2300 stars. 
The total number of cluster stars in \textit{Gaia} DR2 is usually higher than the number of stars used by us, however, 
adding more stars would not have increased the accuracy of our solution, since, as shown in Fig. 1 of \citet{vasiliev2018}, when more than 100 stars are used
to derive the mean proper motion, the formal accuracy of the mean proper motion drops below the large-scale systematic errors of \textit{Gaia} DR2.

We also calculated mean line-of-sight velocities and their errors from the cluster members. For clusters that were not studied by \citet{baumgardt2017} or \citet{bh18}, we first searched the
published literature as well as the ESO and Keck Science archives for additional line-of-sight velocity measurements and unpublished spectra. Additional literature that was used in this paper is listed
in Table \ref{tab:litdata} and the reduction of the archival ESO and Keck data was done as described in \citet{bh18}. We list the mean line-of-sight velocities together with their errors in
Table~\ref{tab:meanpm}. Our solution for the mean line-of-sight velocity of
Ter~9 is somewhat uncertain due to the large number of background stars in the field and the fact that the mean cluster motion is based on only two stars that have both a measured line-of-sight 
velocity and a proper motion in \textit{Gaia} DR2. Our solution for the mean line-of-sight velocity ($29.31 \pm 2.96$ km/sec) also differs significantly from the one found by \citet{vasquezetal2018} ($71.4 \pm 0.4$) km/sec
and \citet{harris1996} ($59 \pm 10$ km/sec). The orbits of 2MASS-GC02 and Mercer 5 are also somewhat uncertain due to the very large extinction of both clusters, which makes identifying members
in \textit{Gaia} difficult. In addition, the line-of-sight velocities of both clusters are uncertain. If our space motion is correct, both clusters move on orbits that are almost perfectly aligned with the Galactic plane and stay within
$\pm$ 200 pc of the plane of the Milky Way's disc.

\subsection{Internal cluster kinematics}

We calculated internal velocity dispersion profiles of globular clusters from both the line-of-sight velocities and the \textit{Gaia} proper motions. For the line-of-sight velocity dispersion profiles, we followed the 
maximum-likelihood approach described by \citet{baumgardt2017} and \citet{bh18}. For stars with measured proper motions in \textit{Gaia} DR2, we used the proper motions to remove non-members before calculating 
the line-of-sight velocity dispersion profiles. For stars without measured proper motions, we used only the line-of-sight velocities to remove non-members. Proper motion cleaning was most useful for the outer parts of bulge 
globular clusters, where it generally lead to a reduction of the velocity dispersion profile and a better agreement with the best-fitting $N$-body models.
In total we were able to determine line-of-sight velocity dispersion profiles for 127 globular clusters.

The proper motion dispersion profiles were calculated from the member stars with accurate proper motions selected in the previous section. We sorted these stars as a function of their distance
from the cluster centre, transformed the proper motion of each star into a coordinate system in which both proper motion components are uncorrelated with each other, and then treated both proper motion
components as independent measurements. The proper motion velocity dispersion profiles were then calculated in the same way as the line-of-sight velocity dispersion profiles. We did not correct for
perspective effects \citep[e.g.][]{vandevenetal2006} since these are usually less than 0.1 km/sec, i.e. significantly smaller than the cluster velocity dispersions themselves. 
In total we could determine internal velocity dispersion profiles for 103 globular clusters
based on the \textit{Gaia} proper motions. For 93 of these we were able to also determine the line-of-sight velocity dispersion profiles, while the remaining 10 only have proper motion dispersion profiles.
Table~\ref{tab:pmdis} lists the proper motion dispersion profiles that we determined from the \textit{Gaia} 
proper motions\footnote{The velocity dispersion profiles together with all other data and figures depicting our fits to individual clusters can also be found at \href{https://people.smp.uq.edu.au/HolgerBaumgardt/globular/}{https://people.smp.uq.edu.au/HolgerBaumgardt/globular/}}.

We also checked for the impact of a possible underestimation of the statistical parallax and proper motion errors in \textit{Gaia} DR2 as found by \citet{lindegrenetal2018} and \citet{mignardetal2018} on 
the internal velocity dispersion profiles. To this end we increased the statistical errors by 10\% and rerun all our fits. We found that the velocity dispersion profiles of bright clusters are not
affected by such an underestimation since the velocity dispersions of such clusters are determined mainly by bright stars which have errors smaller than the internal velocity dispersion. For faint clusters,
the derived velocity dispersions can change by up to 10\%, however even then the changes remain usually within the formal error bars of the velocity dispersions.

\subsection{Surface density profiles and stellar mass functions}

The surface density profiles of 118 globular clusters were already compiled by us based on published literature and our own measurements in \citet{bh18}. For the remaining clusters we took the surface density profiles 
either from \cite{trageretal1995} or other literature papers, as listed in
Table~\ref{tab:litdata}.  When several measurements of the surface density profile were available, we combined them to increase the spatial coverage and accuracy of the observed surface density
profile. HP~1, NGC~6522, NGC~6453, NGC~6540, Ter~1, Ter~3 and Ter~4 did not have well determined published surface density profiles, so we determined surface density profiles for these
clusters from the near-infrared photometric data presented by \citet{valentietal2007} and \citet{valentietal2010}.
\begin{figure*}
\begin{center}
\includegraphics[width=\textwidth]{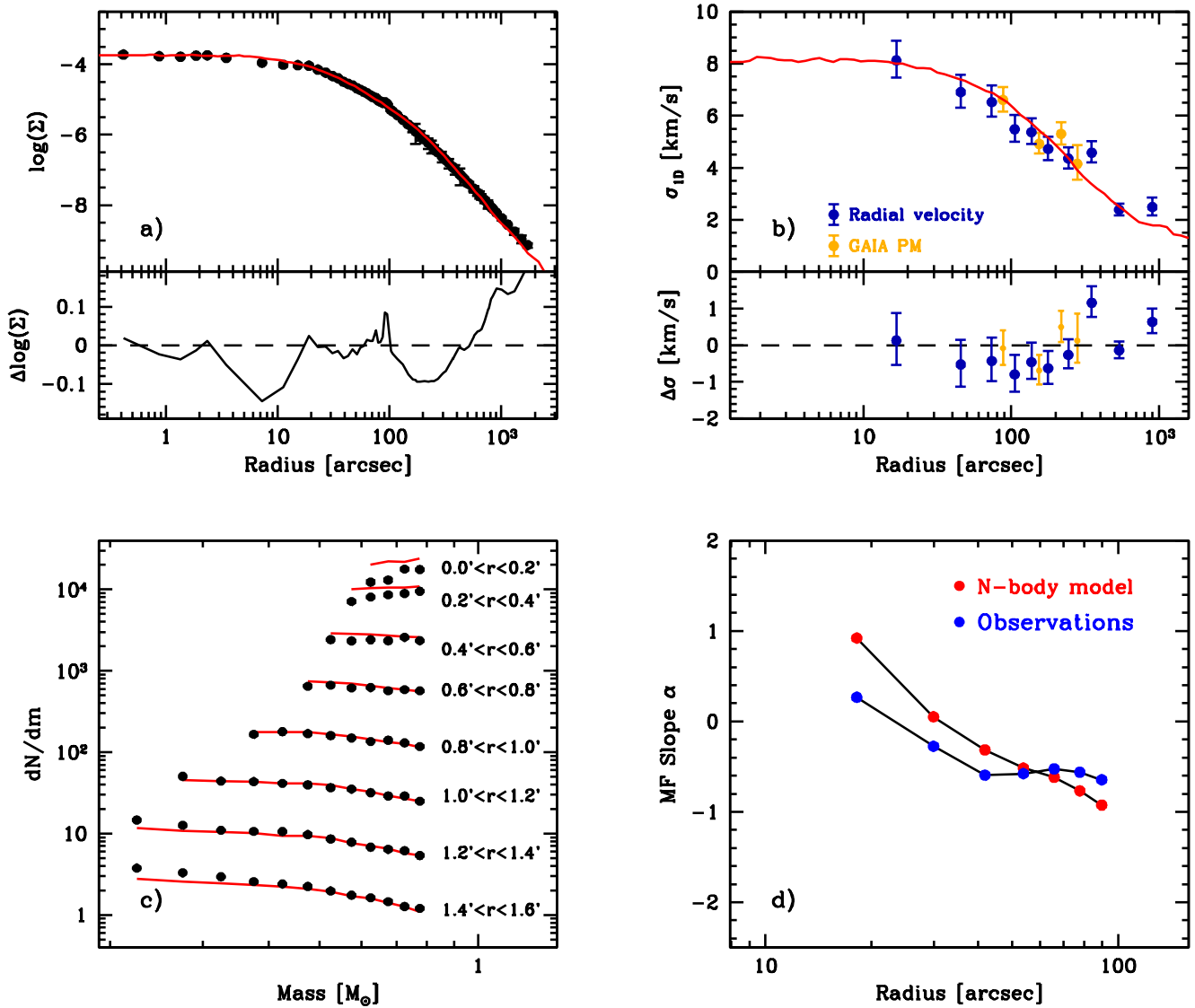}
\end{center}
\caption{Fit of the surface density profile (panel a), velocity dispersion profile (panel b), number of main-sequence stars as a function of stellar mass at 8 different
radii in the cluster (panel c) and mass function slope as a function of radius (panel d) of NGC 5272. In each panel the best-fitting $N$-body model is shown by red lines or dots
while the observed data is shown in other colors. In panel b), the velocity dispersion profile based on the \textit{Gaia} proper motions is shown by orange circles while blue circles
show the line-of-sight velocity dispersion profile. The best-fitting $N$-body model is within 10\% of the observed surface density profile, within 0.5 km/sec of the observed
velocity dispersion profile and within $\Delta \alpha=0.3$ in mass function slope over the whole range of radii. In addition there is very good agreement in the absolute number of
main-sequence stars at different radii between the best-fitting $N$-body model and the observations of NGC 5272.}
\label{fig:ngc5272fit}
\end{figure*}

The stellar mass functions of 42 clusters were already used by us in \citet{bh18}, based mainly on the results obtained by \citet{sollimabaumgardt2017} from an analysis
of the HST/ACS Treasury project data \citep{sarajedinietal2007}. To this data we added stellar mass functions of an additional
55 clusters in order to improve the accuracy of our $N$-body fitting and to be able to fit clusters that do not have kinematic data.
We took 17 of these mass function measurements, based mainly on either HST/WFPC2 or HST/ACS data from published literature and list this data in Table~\ref{tab:litdata}. In addition, we
fitted the HST/ACS photometry of an additional 38 clusters from the MAST archive, using PARSEC isochrones
\citep{bressanetal2012} to transform stellar magnitudes into masses. For the clusters that we fitted ourselves, we corrected the CMDs for variable extinction following the procedures
in \citet{bellinietal2017a} before deriving stellar masses from the isochrones, and applied artificial stars tests according to \citet{bellinietal2017b}
to correct the derived mass functions for incompleteness at the faint end.
 
Together with the mass functions published by \citet{sollimabaumgardt2017}, we now have a sample of 97 globular clusters that have measured mass functions, i.e. nearly 2/3 of all known Galactic globular clusters.
For each cluster we determined the exact location of each individual \textit{HST} field on the sky using the information found in the MAST Archive. For the comparison with the mass functions from the $N$-body models, 
we projected the $N$-body models onto the sky and determined mass functions only for stars that are located within these \textit{HST} fields.  
In total we were able to obtain either kinematic or mass function data for 144 out of the 156 Galactic globular clusters, i.e. about 90\% of all clusters.
\begin{figure*}
\begin{center}
\includegraphics[width=0.9\textwidth]{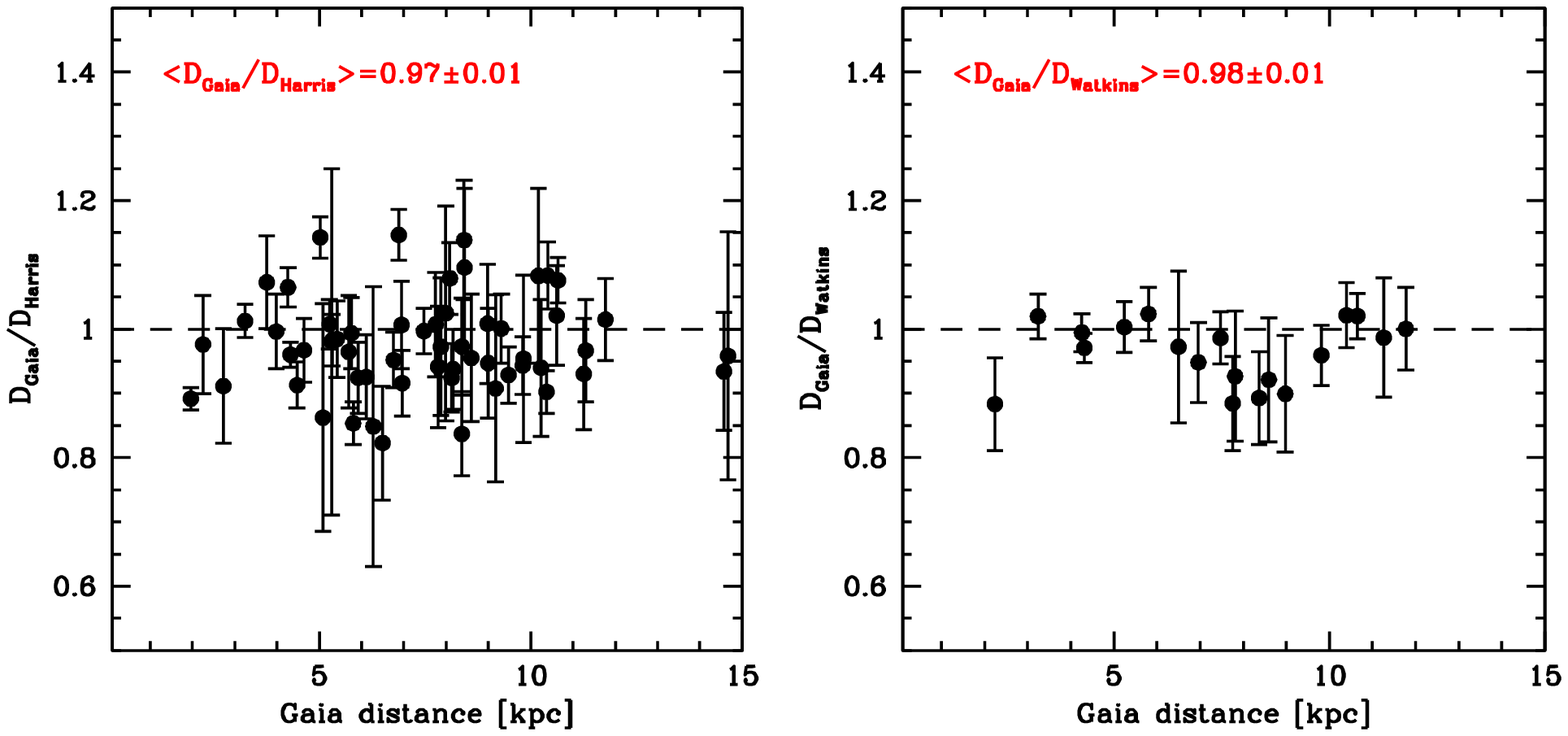}
\end{center}
\caption{Comparison of the best-fitting kinematic cluster distances, derived from fitting the \textit{Gaia} proper motion dispersion profiles together with the line-of-sight velocity dispersion profiles, with
the distances given by \citet{harris1996} (left panel) and the kinematic distances derived by fitting the \textit{HST} proper motion dispersion profiles of \citet{watkinsetal2015a} (right panel). There is excellent 
agreement of the \textit{Gaia} distances with the literature values and the \textit{HST} proper motions, especially for nearby clusters. For more distant clusters the \textit{Gaia} data might lead to slightly
smaller distances than what is derived from the other methods.}
\label{fig:dist1}
\end{figure*}

\subsection{$N$-body fits of the cluster data}

We fitted the line-of-sight velocity and proper motion velocity dispersion profiles calculated in the previous section by a grid of dedicated $N$-body simulations to obtain the cluster parameters like total masses,
half-light and half-mass radii.
The details of the fitting procedure can be found in \cite{baumgardt2017} and \citet{bh18}. In short, we searched a grid of about 1,600 $N$-body simulations of isolated star clusters with varying
sizes, initial density profiles and initial mass functions until we found the model which best reproduces the observed velocity dispersion and surface density profile
and the observed stellar mass function of each globular cluster.
For the fitting, the $N$-body models were scaled to match the size of each individual cluster and we fitted the surface density, velocity dispersion profiles and, if available, the stellar mass function
of the clusters. We interpolated in our grid of $N$-body models to increase the number of models available for comparison and improve the accuracy of our fits.

In this paper we expanded our fitting method to be able to fit clusters that have only a measured mass function but no kinematical information. We did this by guessing a value
for the velocity dispersion at the half-mass radius of the clusters
and then varying this value until our best-fitting $N$-body model predicts the same absolute number and the same distribution of stars over mass as seen in the observed clusters.
We also applied this method to low-mass clusters with kinematic information like Pal~13, since we found that the observed number of cluster stars was significantly below the predicted number based on the velocity
dispersion profile in these clusters. We attribute this difference to the influence of undetected binary stars which inflate the velocity dispersion profile \citep[e.g][]{blechaetal2004} and regard the mass
derived from the mass function fit as the more reliable value.

Fig.~\ref{fig:ngc5272fit} compares as an example the observational data for NGC~5272 (M~3) with our best-fitting $N$-body model. It can be seen that we obtain
a very good fit to the observational data for this cluster. The best-fitting cluster mass that we obtain from matching the observed velocity dispersion profile of NGC~5272 ($M_C=379,000 \pm 19,000$ M$_\odot$) also predicts the correct
number of main sequence stars in the range $0.2<m<0.8$ M$_\odot$ as determined by \citet{sollimabaumgardt2017} in the central 1.6 arcmin. In addition, the \textit{Gaia} proper motion velocity dispersion profile is in good agreement 
with the line-of-sight velocity dispersion profile.
Additional examples of fits to individual globular clusters can be found in \citet{bh18} and on our globular cluster website.

\section{Results}

\subsection{Cluster distances}
\label{sec:distances}

For clusters that have both well determined proper motion and line-of-sight velocity dispersion profiles we can also derive kinematic cluster distances from the fitting of the $N$-body models. To this end we varied
the assumed cluster distance until the combined $\chi^2$ value derived from fitting the $N$-body models to the observed velocity dispersions is minimal. Fig.~\ref{fig:dist1} shows a comparison of the distances
derived in this way against the clusters distances given in the Harris catalogue (left panels) and the kinematic distances derived by fitting our $N$-body models only against the \textit{HST} proper motion dispersion profiles
of \citet{watkinsetal2015a}, which are based on the high-precision \textit{HST} proper motions of \citet{bellinietal2014}. It can be seen that we usually find very good agreement between the distances derived using the \textit{Gaia} proper motions
and either the distances in the Harris catalogue (which are mainly based on isochrone fitting of the clusters' color-magnitude diagrams) and the kinematic distances using only the \textit{HST} data. The average ratio of the \textit{Gaia} distances to the distances
in the Harris catalogue is $D_{Gaia}/D_{Harris} = 0.97 \pm 0.01$ while it is $D_{Gaia}/D_{Watkins} = 0.98 \pm 0.01$ for the kinematic distances based only on \textit{HST} proper motions. Both values are very close to
unity. Also the typical distance differences (8\% vs. the Harris data and 5\% against the Watkins data) are of the same order as what one should expect based on the distance errors.
The agreement is especially
good for distances up to about 7 kpc. For larger distances the \textit{Gaia} distances are on average 10\% smaller than the other distance estimates, which could point to the fact that systematic effects become important
at larger distances due to the small proper motions of the stars.
\begin{figure*}
\begin{center}
\includegraphics[width=0.9\textwidth]{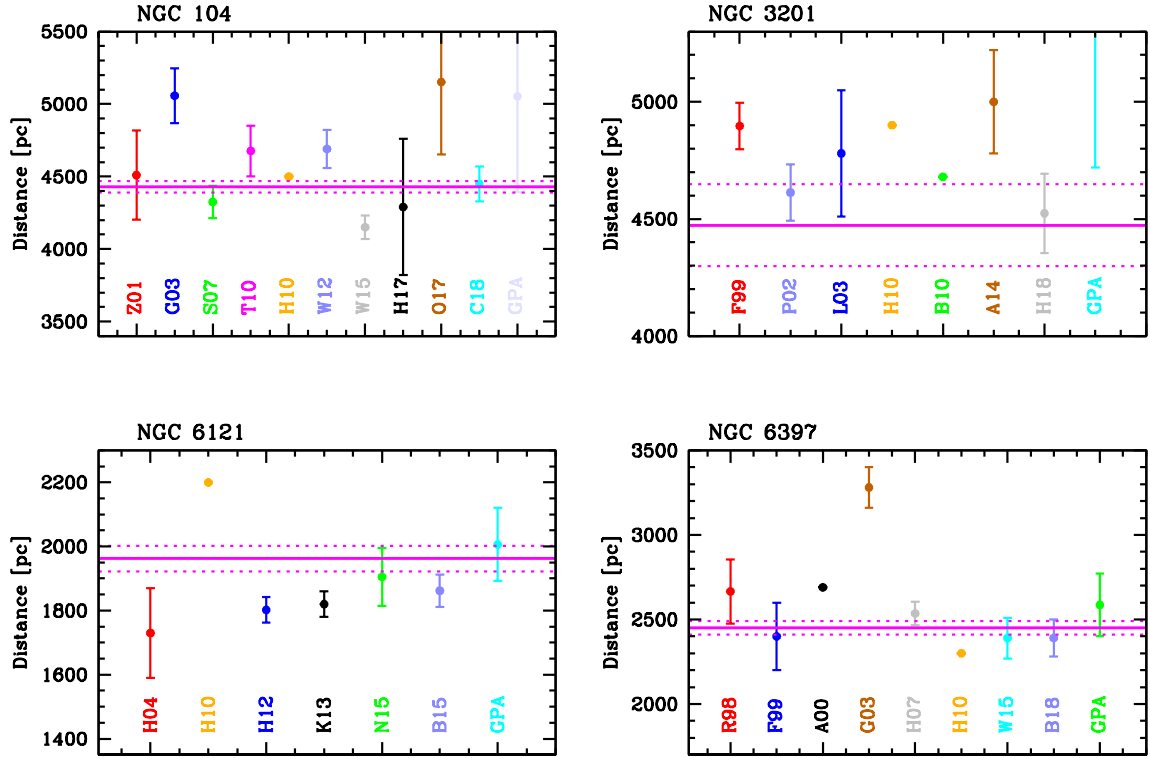}
\end{center}
\caption{Comparison of the best-fitting kinematic cluster distances with
literature estimates for four nearby clusters. The solid and dashed lines show the best-fitting kinematic distance and its $1\sigma$ error bar. GPA denotes the distances derived from averaging the parallaxes
of all \textit{Gaia} members. H10 are the cluster distances given by \citet{harris1996}. All other papers are indicated in the list of references. The kinematic distances agree very well with literature estimates
except for NGC~6121, where our kinematic distance is higher by about 150 pc.}
\label{fig:dist2}
\end{figure*}

Another way to test for the accuracy of the kinematic distances is to compare them for a number of nearby clusters with distance estimates from the literature. We depict such a comparison in Fig.~\ref{fig:dist2}. In this Figure,
the kinematic cluster distance derived from the $N$-body modeling of the cluster is shown by a solid line, with dashed lines indicating the $1\sigma$ upper and lower error bars. Circles indicate a number of recent determinations
from the literature. The abbreviations of the used papers can be found in the list of references. For NGC 104 and NGC 6397 the kinematic distance is derived from a simultaneous fit of the \citet{watkinsetal2015a} and the
\textit{Gaia} proper motions, for the other two clusters it is based solely on the \textit{Gaia} proper motions. It can be seen that for NGC~104, NGC~3201 and NGC~6397, the kinematic distances are in good agreement with 
recent literature determinations. The only significant deviation could be NGC~6121 where our kinematic distance is
about 100 to 150~pc higher than recent literature values. However the \textit{Gaia} distance agrees well with the distance derived from the \textit{Gaia} parallaxes (once a systematic error of 0.04 mas is added to the error of the
mean parallax obtained from the member stars) and the Harris (2010) distance is even larger than the \textit{Gaia} 
distance. We conclude that the kinematic distances that we have determined should be fairly reliable. We therefore calculate new distances to globular clusters by taking a weighted mean of our kinematic distances 
and the distance from the Harris catalogue, assuming a relative distance error of 10\% for the Harris distances. The distances and their $1\sigma$ errors calculated this way are shown by italics in Table~1 and we list
the distances that we obtain from our kinematic fits in Table~\ref{tab:gcdist}.

\subsection{Space orbits and initial mass distribution}

In order to convert the mean proper motions derived in the previous section into velocities, we either used the distances that we determined
kinematically in the previous section, or, for clusters without kinematic distances, we used the distances from \citet{bh18} or \citet{harris1996}.
From the distances and cluster velocities we also derived the space motions of the clusters, assuming a distance of $d=8.1$~kpc 
of the Sun from the Galactic centre \citep{abuteretal2018} and a velocity of (U, V, W)$_\odot$ = (11.1, 12.24, 7.25) km/sec of the Sun relative to the local standard of rest \citep{schoenrichetal2010}. 
We also assumed a circular velocity of 240 km/sec at the position of the Sun, which is in agreement with the proper motion of Sgr A* as determined by
\citet{reidbrunthaler2004} and our adopted distance to the Galactic centre.
We then integrated the cluster orbits backwards in time for 2 Gyrs, using a fourth-order Runge-Kutta integrator and determined the average perigalactic and apogalactic distances
for each cluster. 
Table~\ref{tab:meanpm} lists the values that we obtain for the Milky Way mass model presented in Table~1 of \citet{irrgangetal2013}. Their model is an updated version of the model
suggested by \citet{allensantillan1991} and yields a very good match to recent determinations of the rotation curve of the Milky Way, the in-plane proper motion of Sgr A*, 
and the local mass/surface density.  As a check of the robustness of our results, we also calculated cluster orbits using the Milky Way models of 
\citet{jsh1995} and the best-fitting model of \citet{mcmillan2017}, but found only little difference in the final results compared to the Irrgang et al. model.

In order to get an idea of the initial cluster masses, we also calculated the cluster orbits backward in time over 12 Gyr. In this case we applied dynamical friction to the
cluster orbits according to eq. 7-17 of Binney \& Tremaine (1987):
\begin{equation}
\frac{d \vec{v}}{dt} = -\frac{4 \pi \ln \Lambda G^2 M_{GC} \rho}{v^3} \left[ erf(X)-\frac{2X}{\sqrt{\pi}e^{-X^2}} \right] \vec{v} 
\end{equation}
where $\vec{v}$ is the velocity of a globular cluster, $M_{GC}$ its mass, $\ln \Lambda=10$ the Coulomb logarithm, $\rho$ the density of background stars at the position of the cluster and $X=v/\sqrt{2.0}\sigma$ the ratio
of the globular cluster velocity over the velocity dispersion $\sigma$ of the background stars (taken to be $1/\sqrt{3}$ times the local circular velocity) at the position of the cluster.
During the integration we also applied mass loss to the cluster orbits assuming that clusters lose mass linearly over their lifetime, i.e.:
\begin{equation}
 \frac{d M}{dt} = -\frac{1}{T_{Diss}} M(t)
\end{equation}
The lifetime $T_{Diss}$ of a cluster was calculated based on eqs. 10 and 12 of \citep{baumgardtmakino2003}:
\begin{eqnarray}
\nonumber \frac{T_{Diss}}{\mbox{[Myr]}} & = & 1.35 \left( \frac{M_{Ini}}{\ln(0.02 N_{Ini})} \right)^{0.75}
  \cdot \frac{R_{Apo}}{\mbox{[kpc]}} \cdot \left( \frac{V_G}{240\; \mbox{km/sec}} \right)^{-1} \\
   & & \cdot \left( 1-\epsilon \right)
\end{eqnarray}
and
\begin{equation}
 M(t)= 0.50 M_{Ini} \left( 1-t/T_{Diss} \right) .
\end{equation}
In the above equations, $M_{Ini}$ is the initial mass of a globular cluster, $N_{Ini}=M_{Ini}/<\!\!m\!\!>_{Ini}$, the initial number of cluster stars, $<\!\!m\!\!>_{Ini}=0.65$ M$_\odot$ is 
the initial mean mass of a star in a cluster, $V_G$ is the circular velocity of the Milky Way and $\epsilon$ is the orbital
eccentricity. In the second equation the factor 0.50 reflects the mass loss due to stellar evolution, which reduces the mass of a cluster by about 50\% within the first Gyr of its
evolution.  We note that the above equations were derived
in a spherically symmetric, isothermal galaxy potential and will therefore only give an approximation to the evolution of Milky Way globular clusters.
For each cluster we iterate over the initial mass until we obtain the current mass at the age of each individual cluster. 
\begin{figure}
\begin{center}
\includegraphics[width=\columnwidth]{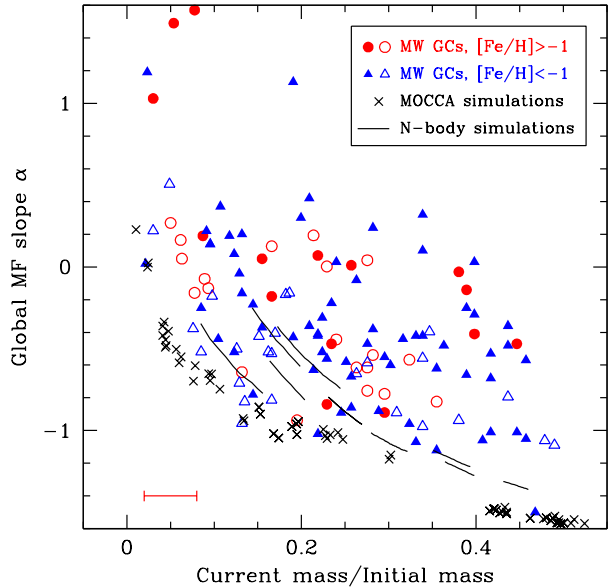}
\end{center}
\caption{Mass fraction $M(t)/M_{ini}$ remaining in a globular cluster vs. the slope of the best-fitting power-law to the mass function of main-sequence stars between 0.2 to 0.8 M$_\odot$. Milky
Way globular clusters are shown by circles and triangles depending on their metallicity, results from $N$-body simulations by solid lines and results from Monte Carlo simulations by crosses. 
Globular clusters with direct mass function
measurements are shown by filled circles/triangles, while clusters where the mass function slope is predicted based on the current relaxation time are shown by open circles and triangles. Due 
to mass segregation and the preferential
loss of low-mass stars, simulated star clusters show a clear correlation between the relative mass loss and the mass function slope of the remaining stars. Observed globular clusters show a similar
correlation but it is shifted by about 0.6 dex towards higher values of the mass function slope, i.e. fewer low-mass stars. The distribution of metal-poor clusters is similar to that of
metal-rich ones, arguing against a mass function varying with metallicity. The errorbar in the lower left corner depicts the typical uncertainty
in $M(t)/M_{ini}$ due to uncertainties in the cluster orbits.}
\label{fig:mfslope}
\end{figure}

Fig.~\ref{fig:mfslope} plots the mass function slope $\alpha$ of the best fitting power-law mass function $N(m) \sim m^{\alpha}$ vs. the relative mass $M(t)/M_{Ini}$ still remaining in the clusters. The mass function slopes
$\alpha$ are derived from the $N$-body fits to the observed mass functions and are valid for main-sequence stars in the mass range $0.2<m<0.8$ M$_\odot$. We splitted the
cluster population into metal-rich clusters with [Fe/H]$>-1$ and metal-poor ones with [Fe/H]$<-1$. Clusters that have measured
mass functions are depicted by filled circles/triangles in this plot while open circles/triangles depict clusters without direct mass determinations. For these clusters the mass functions
were estimated from the relation between half-mass relaxation time $T_{RH}$ and $\alpha$ found by \citet{bh18}: $\alpha = 8.23 - 0.95 \log_{10} T_{RH}$. Here the relaxation time 
$T_{RH}$ of each cluster is also derived from the $N$-body fits.

It can be seen that Galactic globular clusters show a correlation between their present-day mass function slopes $\alpha$ and the amount of mass that has been lost from the clusters, the average 
mass function slope changes from $\alpha=-0.46 \pm 0.11$ for clusters that still retain an average of 40\% of their initial mass to $\alpha=-0.16 \pm 0.15$ for clusters that with 20\% of their initial mass 
to  $\alpha=0.20 \pm 0.17$ for clusters that have inly 10\% of their initial mass remaining. Metal-rich and metal-poor GCs follow more or less the same trend, indicating that the correlation is not driven by variations of the mass
function with metallicity, but must be due to dynamical mass loss, which pushes low-mass stars towards the outer cluster parts where they are easily removed by the tidal field. 
The resulting mass loss decreases the cluster masses and also changes the mass function at the low-mass end towards less negative $\alpha$ values (fewer low-mass stars).
This is confirmed by the solid lines and asterisks, which show the
evolution of star clusters from the $N$-body simulations by \citet{baumgardtsollima2017} and the Monte Carlo simulations by \citet{askaretal2017} respectively.
The data from the simulated clusters is taken at times around $T=12$ Gyr to allow for a direct comparison with the globular clusters. The clusters in the $N$-body simulations start
from Kroupa (2001) mass functions initially while the clusters in the Monte Carlo simulations have a slightly more bottom heavy IMF, which explains their initially more negative $\alpha$ values.
The simulated clusters share the same correlation between mass function slope and mass lost that is seen for the globular clusters, a weaker change in the initial stages followed by stronger
changes close to dissolution.


Globular clusters that have lost only a small portion of their mass through dynamical evolution, have an average mass function 
slope around $\alpha=-0.6$, flatter by about one dex then the best-fitting power-law mass function slope of a Kroupa (2001) mass function for stars in the mass range $0.2 < m < 0.8$ M$_\odot$. 
We take this as strong indication that globular clusters formed with initially bottom-light mass functions. 
Finally, there is very good agreement in the distribution of both clusters with and without
direct mass function determinations, showing that the correlation between the mass lost from a cluster and the overall mass function slope is not driven by selection effects in the cluster sample that has
direct mass function determinations. 
\begin{figure}
\begin{center}
\includegraphics[width=\columnwidth]{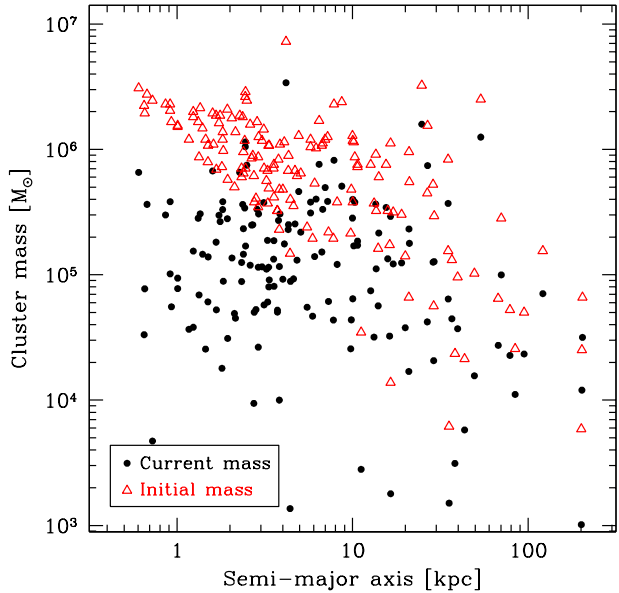}
\end{center}
\caption{Current and initial cluster masses as a function of the semi-major axis of a cluster's orbit in the Milky Way. Nearly all globular clusters with semi-major axes less than 2~kpc have formed with an
initial mass $M>10^6 M_\odot$. If the initial cluster mass function was independent of galactocentric distance, most clusters in the central few kpc of the Milky Way have been destroyed by dynamical evolution.}
\label{fig:mini}
\end{figure}

\subsection{Initial globular cluster population of the Milky Way}

Fig.~\ref{fig:mini} depicts the distribution of the current and initial cluster masses as a function of the present-day semi-major axis $R_{Semi}$ of their orbit in the Milky Way. We calculated the semi-major axes
as the mean of the average apogalactic and perigalactic distance of the cluster orbits. The present-day cluster masses are shown by filled circles. They show a weak correlation with the semi-major
axis of the orbit, a linear relation between $\log M$ and $\log R_{Semi}$ gives as best-fitting relation $\log M = 5.23 - \left(0.30 \pm 0.007\right) \log R_{Semi}$. However most of this trend
is due to clusters beyond $R_{Semi}>30$~kpc, which are on average less massive than the inner clusters. Excluding these, the relation changes to $\log M = 5.09 - \left(0.003 \pm 0.013\right) \log R_{Semi}$
and is compatible with a mass distribution independent of Galactocentric distance. The reason
for the lower masses of outer clusters could either be dynamical evolution having destroyed the $M<10^4$ M$_\odot$, low-mass clusters in the inner galaxy, or point to a different origin 
and formation mode of the outer clusters.

In contrast, the initial masses of the globular clusters which have survived to the present time are strongly increasing with decreasing distance to the Galactic centre.
Inside 2~kpc in particular, most clusters that can still be found in the Milky Way have started with masses larger than $10^6$ M$_\odot$.
This indicates that the current cluster population in the inner parts is most likely just a small portion of the initial population, with the other clusters having either dissolved by the strong
tidal field or spiraled into the centre of the Milky Way and merged with the nuclear cluster of the Milky Way \citep{antoninietal2012}.
We find that the current mass of the Galactic globular cluster system is $3.4 \cdot 10^7$ M$_\odot$, about 1/3 of this mass being in the 10 most massive clusters alone
(M19, NGC~5824, M14, 47~Tuc, NGC~2808, NGC~6338, NGC~6441, NGC~2419, M54 and Omega Cen). The initial mass of the currently surviving clusters was $1.5 \cdot 10^8$ M$_\odot$, i.e. the currently
surviving globular clusters have on average lost nearly 80\% of their mass since the time of their formation through either stellar evolution mass loss or dynamical mass loss. Most of 
the dynamical mass loss is due to mass lost from low mass clusters in the inner parts of the
Milky Way. The average mass-loss reduces slightly to about 75\% for clusters with present-day masses larger then $10^5$ M$_\odot$. This is in agreement with the factor 2 to 4

\begin{figure}
\begin{center}
\includegraphics[width=\columnwidth]{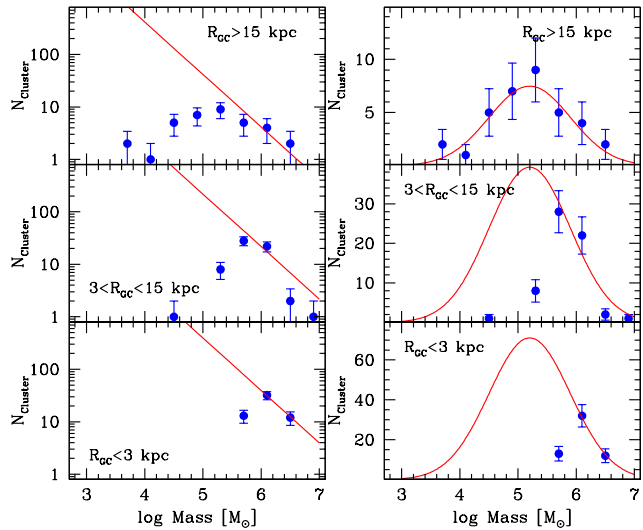}
\end{center}
\caption{Fit of the initial mass distribution of globular clusters with $N(m) \sim m^{-2}$ power-law distributions (left panels) and log-normal distributions (right panels) for clusters with orbital semi-major 
axes larger than 15 kpc (top panels), 3 to 15 kpc (middle panels) and inside 3 kpc (bottom panels). In all panels the theoretical distributions have been matched to the number of globular clusters
with initial masses larger than $10^6$ M$_\odot$. Both types of theoretical distributions provide a good match to the observed cluster distribution for the more massive clusters and imply significant
cluster destruction in the inner parts.}
\label{fig:gausspl}
\end{figure}

In order to obtain a better estimate of how large the dissolved cluster population could have been, we plot the distribution of initial masses at different Galactocentric distances 
and fit log-normal distributions to the massive end of the resulting mass distributions. The initial mass distribution of the outermost clusters in the Milky Way with $R_{Semi}>15$~kpc, 
which should have been least affected by cluster dissolution, is well fitted by a 
log-normal distribution with mean $\log M_{Ini}= 5.2$ and scatter $\sigma_{Ini}=0.7$ (see Fig.~\ref{fig:gausspl}). Fig.~\ref{fig:gausspl} shows the distribution of initial cluster masses
if we split the cluster population into three different radial bins. In each bin, the height of the log-normal distributions is chosen to match the number of clusters with masses 
$M_{Ini}>10^6$ M$_\odot$. This way, we find that the initial number of clusters was 500 with a combined mass of $M_{Tot}=2.5 \cdot 10^8$ M$_\odot$, larger by 50\%
than what is being accounted for by the remaining clusters. We can also fit the cluster distribution
by a power-law mass distribution $N(m) \sim M_{GC}^\beta$, with $\beta=-2.0$, similar to that seen for young star clusters in the Milky Way and other nearby galaxies \citep{hunteretal2003,gielesetal2006b,baumgardtetal2013}. 
Assuming lower and upper cut-off values of $10^4$ M$_\odot$ and $10^7$ M$_\odot$ respectively, we then find that about 7000 globular clusters formed in the Milky Way with a combined
mass of about  $M_{Tot}=5.4 \cdot 10^8$ M$_\odot$, i.e. about a factor two larger than if clusters started with a log-normal mass distribution. The total mass of the Galactic bulge and
halo is however still an order of magnitude higher \citep{sofue2013,irrgangetal2013,blandhawthorngerhard2016}. It has been suggested that a large fraction of the Milky Way bulge
stars have formed through disc instability processes and are not part of a classical bulge formed through mergers early in the formation of the Galaxy \citep[e.g.][]{wegggerhard2013}.
However even if only 25\% of the bulge are part of this classical bulge \citep{shenetal2010}, disrupted globular clusters would still account for only a small fraction of the 
stars in the (classical) bulge and halo, unless most mass was in very low-mass clusters.

\section{Conclusions}

We have derived the mean proper motions and space velocities of 154 Milky Way globular clusters through a combination of the \textit{Gaia} DR2 proper motions and ground-based line-of-sight velocities of individual member stars.
Our mean proper motions show good agreement with the proper motions determined by \citet{helmietal2018} and \citet{vasiliev2018} based on \textit{Gaia} DR2 proper motions. They are also in good agreement
with the proper motions derived by \citet{sohnetal2018} from \textit{HST} data if we take a systematic error of 0.10 mas/yr in the \textit{Gaia} proper motions into account.
For about half of all globular clusters, the space velocities are accurate to a few km/sec, however the errors grow to nearly 100 km/sec for the most distant halo clusters.  The limiting factor on the accuracy 
are currently the systematic errors in the \textit{Gaia} proper motions and uncertainties in the cluster distances. We expect that both these error sources will decrease with future \textit{Gaia} data releases.

We also derived the velocity dispersion profiles of 103 globular clusters based on the \textit{Gaia} proper motions. Together with the line-of-sight velocity dispersion profiles published by \citet{bh18} and new
determinations of the stellar mass function of stars in the clusters, this has allowed us to model the internal kinematics of 144 globular clusters, i.e. more than 90\% of the total cluster
population of the Milky Way through a comparison with the results of a large set of dedicated $N$-body simulations. We have made the derived masses and structural parameters as well as the
velocity dispersion profiles available on our globular cluster website {\footnotesize \url{https://people.smp.uq.edu.au/HolgerBaumgardt/globular/}}.

In order to derive the initial mass and the amount of mass lost from each cluster we integrated the cluster orbits backward in time and applied suitable recipes to account for the effects of 
dynamical friction and mass loss of stars to the clusters. We find a correlation between
the present-day mass function slope and the amount of mass that has been lost from a globular cluster, showing that mass segregation and preferential loss of low-mass stars are important mechanisms shaping 
the mass function of stars in clusters. The mass function slopes
are the same for both metal-rich and metal-poor clusters, arguing against a variation of the stellar mass function with metallicity in globular clusters. However given the significant 
cluster-to-cluster scatter in the mass function slopes, small metallicity dependent mass function variations cannot be ruled out either. 

The dynamically least evolved globular clusters have power-law mass function slopes of $\alpha=-0.6$ for main sequence stars in the range 0.2 to 0.8 M$_\odot$, higher by about one dex than a Kroupa mass function over the 
same mass range. We take this as strong indication that globular clusters have started with bottom-light mass functions. Such a bottom-light mass function has recently been predicted for high red-shift galaxies
with $z>6$ due to heating from the cosmic microwave background radiation by \citet{jermynetal2018}. A bottom-light mass function could ease the tension between the large fraction of second generation stars seen in globular 
clusters and the much smaller fractions expected based on the yields of massive stars \citep{bastianlardo2015,renzinietal2015}, since with fewer low-mass stars present initially, a larger fraction can be polluted.
A detailed numerical modeling of the different formation scenarios suggested for 2nd generation stars and their yields will be necessary in order to see if the mass budget problem can be solved this way. 

Finally, the combined initial mass of all globular clusters that have survived to the present time was around $1.5 \cdot 10^8$~M$_\odot$, a factor 5 larger than their current combined mass. Most surviving clusters in the inner parts
of the Milky Way started with masses larger than $10^6$ M$_\odot$, even though for some clusters their present-day masses are up to a factor 100 lower. For a universal initial cluster mass function independent of galactocentric
radius, this implies a large population of destroyed globular clusters in the inner parts of the Milky Way. The exact number of surviving clusters depends on the initial cluster mass function. If this mass function was
a log-normal mass function, the initial number was around 500 clusters, rising to about 7,000 clusters for a $N(m) \sim m^{-2}$ power-law mass function.

\section*{Acknowledgments}

We thank Roger E. Cohen and Daniel Weisz for sharing their data with us and an anonymous referee for comments that helped improve the presentation of our paper.
This work has made use of data from the European Space Agency (ESA) mission
{\it Gaia} (\url{https://www.cosmos.esa.int/gaia}), processed by the {\it Gaia}
Data Processing and Analysis Consortium (DPAC,
\url{https://www.cosmos.esa.int/web/gaia/dpac/consortium}). Funding for the DPAC
has been provided by national institutions, in particular the institutions
participating in the {\it Gaia} Multilateral Agreement. This paper is also based on data obtained from the ESO Science Archive Facility. This research has made use of the Keck Observatory Archive (KOA), which is operated by the W. M. Keck Observatory and the NASA Exoplanet Science Institute (NExScI), under contract with the National Aeronautics and Space Administration.  Some of the data presented in this paper was obtained from the Mikulski Archive for Space Telescopes (MAST). STScI is operated by the Association 
of Universities for Research in Astronomy, Inc., under NASA contract NAS5-26555. Support for MAST for non-HST data is provided by the NASA Office of Space Science via grant NNX09AF08G 
and by other grants and contracts.

\nocite{zoccalietal2001}
\nocite{grattonetal2003}
\nocite{salarisetal2007}
\nocite{thompsonetal2010}
\nocite{woodleyetal2012}
\nocite{watkinsetal2015b}
\nocite{heyletal2017}
\nocite{omalleyetal2017}
\nocite{chenetal2018}
\nocite{ferraroetal1999}
\nocite{piersimonietal2002}
\nocite{layden2003}
\nocite{bonoetal2010}
\nocite{arellano2014}
\nocite{hansenetal2004}
\nocite{hendricksetal2012}
\nocite{kaluznyetal2013}
\nocite{neeleyetal2015}
\nocite{bragaetal2015}
\nocite{reid1998}
\nocite{anthonytwarog2000}
\nocite{brown2018}
\nocite{hansen2007}

\bibliographystyle{mn2e}
\bibliography{mybib}

\input{spacemotion_tab.tex}

\input{sources.tex}

\input{distances_tab.tex}

\input{velpm_tab.tex}

\label{lastpage}

\end{document}

%% file: spacemotion_tab.tex
\begin{table*}
\caption{Mean proper motions and orbital parameters of 154 Galactic globular clusters. For each cluster, the Table gives the name of the cluster, the right ascension and declination, the distance from the Sun, the mean radial velocity of the cluster and its $1\sigma$ error, the proper motion in right ascension and declination and the correlation $\rho$ between both parameters the X, Y and Z component of the cluster position, and the U, V, W components of the cluster velocity together with their errors. X and U point from the Galactic centre towards the Sun, Y and V point in the direction of Galactic rotation at the Solar position, and Z and W point towards the North Galactic pole. The velocities have been corrected for Galactic rotation and solar motion relative to the LSR. Distances shown in italics are the weighted mean of our kinematic distances and the distances from Harris (2010).}

\begin{tabular}{lr@{}r@{$\;$}r@{$\;$}c@{$\!\!\!$}c@{}c@{$\;\;$}r@{$\;$}r@{$\;$}r@{$\;$}r@{$\;$}r@{$\;$}r@{$\;$}r@{$\;$}r@{$\;$}r}
\hline
 & & & & & & & & & & & & & \\[-0.3cm]
\multirow{2}{*}{Name} &  \multicolumn{1}{c}{RA} & \multicolumn{1}{c}{$R_\odot$} & \multicolumn{1}{c}{$R_V$} & \multicolumn{1}{c}{$\mu_{\alpha*}$} & \multicolumn{1}{c}{$\mu_\delta$} & $\rho_{\mu_\alpha \mu_\delta}$ & \multicolumn{1}{c}{X} & \multicolumn{1}{c}{Y} & \multicolumn{1}{c}{Z} & \multicolumn{1}{c}{U} & \multicolumn{1}{c}{V} & \multicolumn{1}{c}{W} & $R_{Per}$ & $R_{Apo}$ \\
 &  \multicolumn{1}{c}{DEC} & \multicolumn{1}{c}{$\sigma_R$} & \multicolumn{1}{c}{$\sigma_V$} & \multicolumn{1}{c}{$\sigma_{\alpha*}$} & \multicolumn{1}{c}{$\sigma_\delta$} & & \multicolumn{1}{c}{$\sigma_X$} & \multicolumn{1}{c}{$\sigma_Y$} & \multicolumn{1}{c}{$\sigma_Z$} & \multicolumn{1}{c}{$\sigma_U$} & \multicolumn{1}{c}{$\sigma_V$} & \multicolumn{1}{c}{$\sigma_W$} & $\sigma_{Per}$ & $\sigma_{Apo}$ \\
 & \multicolumn{1}{c}{[deg]} & \multicolumn{1}{c}{[kpc]} & \multicolumn{1}{c}{[km/s]} & \multicolumn{1}{c}{[mas/yr]} & \multicolumn{1}{c}{[mas/yr]} & & [kpc] & [kpc] & [kpc] & [km/s] &  [km/s] & [km/s] & [kpc] & kpc] \\[+0.1cm]
\hline
 & & & & & & & & & & & &  & \\[-0.3cm]
NGC 104 &   6.023625 & {\it    4.43} &  -17.21 & $\,$    5.25 &     -2.53 & $\,$    0.04 &    6.26 &   -2.54  &   -3.13 &   75.93 &  172.96 &   44.89 &    5.46 &    7.44 \\
   & -72.081276 & {\it   0.04} &    0.18 & $\,$   0.01 & $\,$   0.01 & &    0.02 &    0.02 &    0.03 &    0.26 &    0.24 &    0.23 &    0.01 &    0.01 \\[+0.05cm]
NGC 288 &  13.188500 & {\it    9.98} &  -44.83 & $\,$    4.22 &     -5.65 & $\,$    0.31 &    8.19 &    0.05  &   -9.98 &    9.89 &  -80.69 &   50.55 &    3.33 &   13.01 \\
\hline
\end{tabular}
\label{tab:meanpm}
\end{table*}

\begin{table*}
\contcaption{}
\begin{tabular}{lr@{}r@{$\;$}r@{$\;$}c@{$\!\!\!$}c@{}c@{$\;\;$}r@{$\;$}r@{$\;$}r@{$\;$}r@{$\;$}r@{$\;$}r@{$\;$}r@{$\;$}r@{$\;$}r}
\hline
 & & & & & & & & & & & & & \\[-0.3cm]
\multirow{2}{*}{Name} &  \multicolumn{1}{c}{RA} & \multicolumn{1}{c}{$R_\odot$} & \multicolumn{1}{c}{$R_V$} & \multicolumn{1}{c}{$\mu_{\alpha*}$} & \multicolumn{1}{c}{$\mu_\delta$} & $\rho_{\mu_\alpha \mu_\delta}$ & \multicolumn{1}{c}{X} & \multicolumn{1}{c}{Y} & \multicolumn{1}{c}{Z} & \multicolumn{1}{c}{U} & \multicolumn{1}{c}{V} & \multicolumn{1}{c}{W} & $R_{Per}$ & $R_{Apo}$ \\
 &  \multicolumn{1}{c}{DEC} & \multicolumn{1}{c}{$\sigma_R$} & \multicolumn{1}{c}{$\sigma_V$} & \multicolumn{1}{c}{$\sigma_{\alpha*}$} & \multicolumn{1}{c}{$\sigma_\delta$} & & \multicolumn{1}{c}{$\sigma_X$} & \multicolumn{1}{c}{$\sigma_Y$} & \multicolumn{1}{c}{$\sigma_Z$} & \multicolumn{1}{c}{$\sigma_U$} & \multicolumn{1}{c}{$\sigma_V$} & \multicolumn{1}{c}{$\sigma_W$} & $\sigma_{Per}$ & $\sigma_{Apo}$ \\
 & \multicolumn{1}{c}{[deg]} & \multicolumn{1}{c}{[kpc]} & \multicolumn{1}{c}{[km/s]} & \multicolumn{1}{c}{[mas/yr]} & \multicolumn{1}{c}{[mas/yr]} & & [kpc] & [kpc] & [kpc] & [km/s] &  [km/s] & [km/s] & [kpc] & kpc] \\[+0.1cm]
\hline
 & & & & & & & & & & & &  & \\[-0.3cm]
NGC 5139 & 201.696838 & {\it    5.24} &  234.28 &     -3.24 &     -6.73 &     -0.47 &    4.91 &   -3.93  &    1.35 &  -98.73 &  -19.45 &  -81.02 &    1.35 &    7.00 \\
   & -47.479584 & {\it   0.07} &    0.24 & $\,$   0.01 & $\,$   0.01 & &    0.04 &    0.05 &    0.02 &    0.32 &    0.30 &    0.36 &    0.04 &    0.04 \\[+0.05cm]
NGC 5272 & 205.548416 & {\it    9.59} & -147.28 &     -0.14 &     -2.64 &     -0.05 &    6.71 &    1.26  &    9.40 &  -60.46 &  135.49 & -134.57 &    5.44 &   15.14 \\
\hline
\end{tabular}
\end{table*}

\begin{table*}
\contcaption{}
\begin{tabular}{lr@{}r@{$\;$}r@{$\;$}c@{$\!\!\!$}c@{}c@{$\;\;$}r@{$\;$}r@{$\;$}r@{$\;$}r@{$\;$}r@{$\;$}r@{$\;$}r@{$\;$}r@{$\;$}r}
\hline
 & & & & & & & & & & & & & \\[-0.3cm]
\multirow{2}{*}{Name} &  \multicolumn{1}{c}{RA} & \multicolumn{1}{c}{$R_\odot$} & \multicolumn{1}{c}{$R_V$} & \multicolumn{1}{c}{$\mu_{\alpha*}$} & \multicolumn{1}{c}{$\mu_\delta$} & $\rho_{\mu_\alpha \mu_\delta}$ & \multicolumn{1}{c}{X} & \multicolumn{1}{c}{Y} & \multicolumn{1}{c}{Z} & \multicolumn{1}{c}{U} & \multicolumn{1}{c}{V} & \multicolumn{1}{c}{W} & $R_{Per}$ & $R_{Apo}$ \\
 &  \multicolumn{1}{c}{DEC} & \multicolumn{1}{c}{$\sigma_R$} & \multicolumn{1}{c}{$\sigma_V$} & \multicolumn{1}{c}{$\sigma_{\alpha*}$} & \multicolumn{1}{c}{$\sigma_\delta$} & & \multicolumn{1}{c}{$\sigma_X$} & \multicolumn{1}{c}{$\sigma_Y$} & \multicolumn{1}{c}{$\sigma_Z$} & \multicolumn{1}{c}{$\sigma_U$} & \multicolumn{1}{c}{$\sigma_V$} & \multicolumn{1}{c}{$\sigma_W$} & $\sigma_{Per}$ & $\sigma_{Apo}$ \\
 & \multicolumn{1}{c}{[deg]} & \multicolumn{1}{c}{[kpc]} & \multicolumn{1}{c}{[km/s]} & \multicolumn{1}{c}{[mas/yr]} & \multicolumn{1}{c}{[mas/yr]} & & [kpc] & [kpc] & [kpc] & [km/s] &  [km/s] & [km/s] & [kpc] & kpc] \\[+0.1cm]
\hline
 & & & & & & & & & & & &  & \\[-0.3cm]
FSR 1735 & 253.044174 &    9.80 &  -69.85 &     -4.40 &     -0.49 &     -0.16 &   -1.06 &   -3.48  &   -0.32 &  102.19 &  137.64 &  152.65 &    0.87 &    5.39 \\
   & -47.058055 &   &    4.88 & $\,$   0.18 & $\,$   0.15 & &    0.92 &    0.35 &    0.03 &    5.33 &    7.44 &    7.98 &    0.27 &    0.83 \\[+0.05cm]
NGC 6235 & 253.355458 &   13.52 &  126.68 &     -3.96 &     -7.61 & $\,$    0.28 &   -5.04 &   -0.25  &    3.16 & -145.64 & -292.77 &  -53.16 &    5.43 &   21.81 \\
\hline
\end{tabular}
\end{table*}

\begin{table*}
\contcaption{}
\begin{tabular}{lr@{}r@{$\;$}r@{$\;$}c@{$\!\!\!$}c@{}c@{$\;\;$}r@{$\;$}r@{$\;$}r@{$\;$}r@{$\;$}r@{$\;$}r@{$\;$}r@{$\;$}r@{$\;$}r}
\hline
 & & & & & & & & & & & & & \\[-0.3cm]
\multirow{2}{*}{Name} &  \multicolumn{1}{c}{RA} & \multicolumn{1}{c}{$R_\odot$} & \multicolumn{1}{c}{$R_V$} & \multicolumn{1}{c}{$\mu_{\alpha*}$} & \multicolumn{1}{c}{$\mu_\delta$} & $\rho_{\mu_\alpha \mu_\delta}$ & \multicolumn{1}{c}{X} & \multicolumn{1}{c}{Y} & \multicolumn{1}{c}{Z} & \multicolumn{1}{c}{U} & \multicolumn{1}{c}{V} & \multicolumn{1}{c}{W} & $R_{Per}$ & $R_{Apo}$ \\
 &  \multicolumn{1}{c}{DEC} & \multicolumn{1}{c}{$\sigma_R$} & \multicolumn{1}{c}{$\sigma_V$} & \multicolumn{1}{c}{$\sigma_{\alpha*}$} & \multicolumn{1}{c}{$\sigma_\delta$} & & \multicolumn{1}{c}{$\sigma_X$} & \multicolumn{1}{c}{$\sigma_Y$} & \multicolumn{1}{c}{$\sigma_Z$} & \multicolumn{1}{c}{$\sigma_U$} & \multicolumn{1}{c}{$\sigma_V$} & \multicolumn{1}{c}{$\sigma_W$} & $\sigma_{Per}$ & $\sigma_{Apo}$ \\
 & \multicolumn{1}{c}{[deg]} & \multicolumn{1}{c}{[kpc]} & \multicolumn{1}{c}{[km/s]} & \multicolumn{1}{c}{[mas/yr]} & \multicolumn{1}{c}{[mas/yr]} & & [kpc] & [kpc] & [kpc] & [km/s] &  [km/s] & [km/s] & [kpc] & kpc] \\[+0.1cm]
\hline
 & & & & & & & & & & & &  & \\[-0.3cm]
NGC 6388 & 264.071777 & {\it   10.74} &   82.85 &     -1.33 &     -2.68 & $\,$    0.12 &   -2.23 &   -2.66  &   -1.26 &  -51.29 &   85.24 &  -16.91 &    1.11 &    3.79 \\
   & -44.735500 & {\it   0.12} &    0.48 & $\,$   0.01 & $\,$   0.01 & &    0.12 &    0.03 &    0.01 &    0.48 &    0.56 &    0.57 &    0.02 &    0.09 \\[+0.05cm]
NGC 6402 & 264.400391 & {\it    9.31} &  -60.71 &     -3.66 &     -5.03 & $\,$    0.24 &   -0.28 &    3.27  &    2.38 &  -47.64 &  -25.80 &   23.42 &    0.65 &    4.35 \\
\hline
\end{tabular}
\end{table*}

\begin{table*}
\contcaption{}
\begin{tabular}{lr@{}r@{$\;$}r@{$\;$}c@{$\!\!\!$}c@{}c@{$\;\;$}r@{$\;$}r@{$\;$}r@{$\;$}r@{$\;$}r@{$\;$}r@{$\;$}r@{$\;$}r@{$\;$}r}
\hline
 & & & & & & & & & & & & & \\[-0.3cm]
\multirow{2}{*}{Name} &  \multicolumn{1}{c}{RA} & \multicolumn{1}{c}{$R_\odot$} & \multicolumn{1}{c}{$R_V$} & \multicolumn{1}{c}{$\mu_{\alpha*}$} & \multicolumn{1}{c}{$\mu_\delta$} & $\rho_{\mu_\alpha \mu_\delta}$ & \multicolumn{1}{c}{X} & \multicolumn{1}{c}{Y} & \multicolumn{1}{c}{Z} & \multicolumn{1}{c}{U} & \multicolumn{1}{c}{V} & \multicolumn{1}{c}{W} & $R_{Per}$ & $R_{Apo}$ \\
 &  \multicolumn{1}{c}{DEC} & \multicolumn{1}{c}{$\sigma_R$} & \multicolumn{1}{c}{$\sigma_V$} & \multicolumn{1}{c}{$\sigma_{\alpha*}$} & \multicolumn{1}{c}{$\sigma_\delta$} & & \multicolumn{1}{c}{$\sigma_X$} & \multicolumn{1}{c}{$\sigma_Y$} & \multicolumn{1}{c}{$\sigma_Z$} & \multicolumn{1}{c}{$\sigma_U$} & \multicolumn{1}{c}{$\sigma_V$} & \multicolumn{1}{c}{$\sigma_W$} & $\sigma_{Per}$ & $\sigma_{Apo}$ \\
 & \multicolumn{1}{c}{[deg]} & \multicolumn{1}{c}{[kpc]} & \multicolumn{1}{c}{[km/s]} & \multicolumn{1}{c}{[mas/yr]} & \multicolumn{1}{c}{[mas/yr]} & & [kpc] & [kpc] & [kpc] & [km/s] &  [km/s] & [km/s] & [kpc] & kpc] \\[+0.1cm]
\hline
 & & & & & & & & & & & &  & \\[-0.3cm]
IC 1276 & 272.684166 & {\it    5.39} &  155.06 &     -2.47 &     -4.41 & $\,$    0.15 &    3.12 &    1.99  &    0.53 & -202.12 &  189.79 &   24.67 &    3.47 &    5.76 \\
   &  -7.207611 & {\it   0.51} &    0.69 & $\,$   0.03 & $\,$   0.03 & &    0.47 &    0.19 &    0.05 &    0.72 &    0.87 &    0.89 &    0.12 &    0.60 \\[+0.05cm]
Ter 12 & 273.065833 &    3.40 &   94.77 &     -6.07 &     -2.63 & $\,$    0.38 &    4.74 &    0.49  &   -0.12 & -119.39 &  183.21 &   69.25 &    2.99 &    5.82 \\
\hline
\end{tabular}
\end{table*}

\begin{table*}
\contcaption{}
\begin{tabular}{lr@{}r@{$\;$}r@{$\;$}c@{$\!\!\!$}c@{}c@{$\;\;$}r@{$\;$}r@{$\;$}r@{$\;$}r@{$\;$}r@{$\;$}r@{$\;$}r@{$\;$}r@{$\;$}r}
\hline
 & & & & & & & & & & & & & \\[-0.3cm]
\multirow{2}{*}{Name} &  \multicolumn{1}{c}{RA} & \multicolumn{1}{c}{$R_\odot$} & \multicolumn{1}{c}{$R_V$} & \multicolumn{1}{c}{$\mu_{\alpha*}$} & \multicolumn{1}{c}{$\mu_\delta$} & $\rho_{\mu_\alpha \mu_\delta}$ & \multicolumn{1}{c}{X} & \multicolumn{1}{c}{Y} & \multicolumn{1}{c}{Z} & \multicolumn{1}{c}{U} & \multicolumn{1}{c}{V} & \multicolumn{1}{c}{W} & $R_{Per}$ & $R_{Apo}$ \\
 &  \multicolumn{1}{c}{DEC} & \multicolumn{1}{c}{$\sigma_R$} & \multicolumn{1}{c}{$\sigma_V$} & \multicolumn{1}{c}{$\sigma_{\alpha*}$} & \multicolumn{1}{c}{$\sigma_\delta$} & & \multicolumn{1}{c}{$\sigma_X$} & \multicolumn{1}{c}{$\sigma_Y$} & \multicolumn{1}{c}{$\sigma_Z$} & \multicolumn{1}{c}{$\sigma_U$} & \multicolumn{1}{c}{$\sigma_V$} & \multicolumn{1}{c}{$\sigma_W$} & $\sigma_{Per}$ & $\sigma_{Apo}$ \\
 & \multicolumn{1}{c}{[deg]} & \multicolumn{1}{c}{[kpc]} & \multicolumn{1}{c}{[km/s]} & \multicolumn{1}{c}{[mas/yr]} & \multicolumn{1}{c}{[mas/yr]} & & [kpc] & [kpc] & [kpc] & [km/s] &  [km/s] & [km/s] & [kpc] & kpc] \\[+0.1cm]
\hline
 & & & & & & & & & & & &  & \\[-0.3cm]
NGC 6838 & 298.443726 & {\it    3.99} &  -22.27 &     -3.41 &     -2.61 & $\,$    0.14 &    5.92 &    3.33  &   -0.32 &  -63.25 &  194.38 &   39.17 &    4.77 &    7.08 \\
   &  18.779194 & {\it   0.20} &    0.19 & $\,$   0.01 & $\,$   0.01 & &    0.11 &    0.17 &    0.02 &    0.23 &    0.21 &    0.24 &    0.06 &    0.01 \\[+0.05cm]
NGC 6864 & 301.519541 &   21.61 & -189.08 &     -0.58 &     -2.78 & $\,$    0.25 &  -10.16 &    6.75  &   -9.39 &   66.92 &  -83.26 &   49.40 &    2.06 &   17.98 \\
\hline
\end{tabular}
\end{table*}

%% file: sources.tex
\begin{table}
\caption{Sources of published individual stellar radial velocities (LOS), surface density profiles (SD)
 and stellar mass functions (MF) used in this work in addition to the data used by \citet{bh18} and \citet{baumgardt2017}.}
\begin{tabular}[h]{c@{\hspace{0.2cm}}c@{\hspace{0.2cm}}c}
\hline
Name & Source & Type \\ 
\hline
2MASS-GC02 & \citet{borissovaetal2007} & LOS \\[+0.1cm]
AM 1     & \citet{dotteretal2008b} & MF \\
         & \citet{hilker2006} & SD \\
         & \citet{suntzeffetal1985} & LOS \\[+0.1cm]
AM 4     & \citet{hamrenetal2013} & MF \\
         & \citet{munozetal2018} & SD \\
         & \citet{carraro2009} & SD \\
\hline
\end{tabular}
\label{tab:litdata}
\end{table}

\begin{table}
\contcaption{}
\begin{tabular}[h]{c@{\hspace{0.2cm}}c@{\hspace{0.2cm}}c}
\hline
Name & Source & Type \\
\hline
Pal 13   & \citet{hamrenetal2013} & MF \\[+0.1cm]
Pal 15   & \citet{pl1989} & LOS \\
         & \citet{munozetal2018} & SD \\[+0.1cm]
Pyxis    & \citet{munozetal2018} & SD \\[+0.1cm]
Ter 4    & \citet{origliarich2004} & LOS \\
         & \citet{bonattobica2008} & SD \\[+0.1cm]
Ter 12   & \citet{cote1999} & LOS \\[+0.1cm]
UKS 1    & \citet{origliaetal2005} & LOS \\[+0.1cm]
Whiting 1& \citet{carraroetal2007} & SD, MF \\ 
         & \citet{munozetal2018} & SD \\[+0.1cm]
\hline
\end{tabular}
\end{table}

%% file: distances_tab.tex
\begin{table}
\caption{Kinematic distances and their 1$\sigma$ errors derived from the fit of our $N$-body models to the proper motion and radial velocity 
dispersion profiles of individual clusters.}
\begin{center}
\begin{tabular}{llrr}
\hline
 \multicolumn{1}{c}{Name} & \multicolumn{1}{c}{Alt.}& \multicolumn{1}{c}{$D_{Kin}$} & \multicolumn{1}{c}{$D_{Harris}$}\\
   & \multicolumn{1}{c}{name} & \multicolumn{1}{c}{[kpc]} & \multicolumn{1}{c}{[kpc]} \\
\hline
IC 1276  &  Pal 7        &  5.29 $\pm$ 1.45 & 5.4$\;\;\;$\\
NGC 104  &  47 Tuc       &  4.45 $\pm$ 0.04 & 4.5$\;\;\;$\\
NGC 288  &               & 10.16 $\pm$ 0.35 & 8.9$\;\;\;$\\ 
NGC 362  &               &  9.22 $\pm$ 0.28 & 8.6$\;\;\;$\\ 
NGC 1851 &               & 11.32 $\pm$ 0.20 & 12.1$\;\;\;$\\ 
NGC 2808 &               & 10.22 $\pm$ 0.12 & 9.6$\;\;\;$\\
\hline
\end{tabular}
\end{center}
\label{tab:gcdist}
\end{table}

%% file: velpm_tab.tex
\begin{table}
\caption{Proper motion dispersion profiles of 103 globular clusters derived from {\textit Gaia} DR2 proper motions. The table gives the name of the cluster, the number of stars used to calculate the 
proper motion dispersion in each bin, the average distance of stars from the cluster centre, and the proper motion dispersion together with the upper and lower $1\sigma$ error bars. Only a fraction of the data is shown
here, the full version of Table~\ref{tab:pmdis} is available online.}
\begin{center}
\begin{tabular}{lcr@{.}lccc}
\hline
 & \\[-0.3cm]
\multirow{2}{*}{Name} & \multirow{2}{*}{$N_{PM}$} & \multicolumn{2}{c}{$r$} & $\sigma$ & \multicolumn{1}{c}{$\Delta \sigma_{up}$} & \multicolumn{1}{c}{$\Delta \sigma_{low}$}\\
 & & \multicolumn{2}{c}{[arcsec]} & \multicolumn{1}{c}{[mas/yr]} & \multicolumn{1}{c}{[mas/yr]} & \multicolumn{1}{c}{[mas/yr]}\\[+0.1cm]
\hline
 & \\[-0.3cm]
BH 261 & 18 & 18 & 19 & 0.144 & 0.034 & 0.025 \\
       & 18 & 48 & 50 & 0.078 & 0.028 & 0.019 \\
       & 15 & 110 & 32 & 0.063 & 0.029 & 0.018 \\
 & \\[-0.3cm]
Djor 2 & 70 & 66 & 93 & 0.071 & 0.043 & 0.071  \\
 & \\[-0.3cm]
ESO 452 & 42 & 50 & 49 & 0.023 & 0.034 & 0.023  \\
 & \\[-0.3cm]
 \multicolumn{1}{c}{$\vdots$} & $\vdots$  & \multicolumn{2}{c}{$\vdots$} & $\vdots$ & $\vdots$ & $\vdots$  \\
\hline
\end{tabular}
\end{center}
\label{tab:pmdis}
\end{table}